\documentstyle[epsf]{mn}

\title[]{{\sl XMM-Newton} and {\sl Chandra} observations of the
ultra-compact binary RX J1914+24}

\author[]
{Gavin Ramsay$^{1}$, Mark Cropper$^{1}$, Pasi Hakala$^{2,3}$\\
$^{1}$Mullard Space Science Laboratory, University College London,
Holmbury St. Mary, Dorking, Surrey, RH5 6NT, UK\\
$^{2}$Observatory, University of Helsinki, PO Box 14,
FIN-00014 University of Helsinki, Finland\\
$^{3}$Tuorla Observatory, V\"ais\"al\"antie 20, 21500 Piikki\"o, Finland\\
}

\begin{document}
\outer\def\gtae {$\buildrel {\lower3pt\hbox{$>$}} \over 
{\lower2pt\hbox{$\sim$}} $}
\outer\def\ltae {$\buildrel {\lower3pt\hbox{$<$}} \over 
{\lower2pt\hbox{$\sim$}} $}
\newcommand{\ergscm} {ergs s$^{-1}$ cm$^{-2}$}
\newcommand{\ergss} {ergs s$^{-1}$}
\newcommand{\ergsd} {ergs s$^{-1}$ $d^{2}_{100}$}
\newcommand{\pcmsq} {cm$^{-2}$}
\newcommand{\ros} {\sl ROSAT}
\newcommand{\asca} {\sl ASCA}
\newcommand{\xmm} {\sl XMM-Newton}
\newcommand{\chan} {\sl Chandra}
\def\rchi{{${\chi}_{\nu}^{2}$}}
\def\uchi{{${\chi}^{2}$}}
\newcommand{\Msun} {$M_{\odot}$}
\newcommand{\Mwd} {$M_{wd}$}
\def\Mdot{\hbox{$\dot M$}}
\def\mdot{\hbox{$\dot m$}}

\maketitle

\begin{abstract}

The nature of the X-ray source RX J1914+24 has been the subject of
much debate. It shows a prominent period of 569 sec in X-rays and the
optical/infra-red: in most models this has been interpreted as the
binary orbital period. We present our analysis of new {\xmm} and
{\chan} data. We find a longer term trend in the {\xmm} data and power
at 556 and 585 sec in 5 sets of data. It is not clear if they are
produced as a result of a beat between a longer intrinsic period and
the 569 sec modulation or if they are due to secular variations. We
obtain a good fit to the {\xmm} spectrum with a low temperature
thermal plasma model with an edge at 0.83keV. This model implies an
unabsorbed bolometric X-ray luminosity of $1\times10^{33}$ \ergss (for
a distance of 1kpc) - this is 2 orders of magnitude lower than our
previous estimate (derived using a different model). If the distance
is much less, as the absorption derived from the X-ray fits suggest,
then it is even lower at $\sim3\times10^{31}$ \ergss.

\end{abstract}


\section{Introduction}

The nature of the X-ray source RX J1914+24 has been the subject of
much debate in recent years. Only one modulation period (569 sec) has
been seen in both its X-ray and optical light curves (Ramsay et al
2000). For this, and other reasons, the period of 569 sec has been
taken to be the binary orbital period, which would make it the second
most compact binary system known.

The models which have been put forward to explain the observational
characteristics of RX J1914+24 fall into two broad categories:
accretion and non-accretion driven models. The fact that the optical
spectrum of RX J1914+24 (also known as V407 Vul) shows no emission
lines, puts a question mark over the accretion driven models. The
non-accreting model is the unipolar-inductor (UI) model of Wu et al
(2002). While this model has been able to explain many of the observed
characteristics of this source, it appears that its X-ray luminosity
implies an induction torque which is much greater than the amount
generated by gravitational radiation. There is some evidence that the
observed spin-up rate of RX J1914+24 (Strohmayer 2002, 2004b, Ramsay et
al 2005a) is therefore much smaller than that predicted by the UI model
(Marsh \& Nelemans 2005). This conclusion is based on the X-ray
luminosity determined using {\ros} (Cropper et al 1998) and {\xmm}
observations (Ramsay et al 2005a). The {\ros} data were well fitted
using an absorbed blackbody model while the higher signal to noise
data of {\xmm} found a poor fit using this model. Ramsay et al (2005a)
found the best fit to the data could be achieved using an absorbed
blackbody plus Gaussian model, although they did not expect that this
model was physically realistic.

Since those first {\xmm} observations were made in Nov 2003, a further
two sets of observations made in Sept/Oct 2004 have become publically
available. These observations were longer than the initial {\xmm}
observations and the particle background was lower. In addition, a
further set of X-ray observations of RX J1914+24 made using {\chan}
has become publically available.  We have therefore extracted these
data from the archive and use them to make a more detailed
investigation of its X-ray spectrum, to determine if the 569 sec is
continuing to spin-up at the same rate and to search for evidence for
periods other than the 569 sec period.

\section{Observations and Data Reduction}

\subsection{{\xmm} observations}

RX J1914+24 was observed twice using {\xmm} in Nov 2003. The details
of these observations has been reported by Ramsay et al (2005a).  It
was observed again late in 2004 and we show in Table \ref{log} the
start time, duration and mean X-ray count rate of the second epoch
{\xmm} observations. In this paper we discuss the data collected using
the EPIC X-ray detectors (pn and MOS) which are sensitive over the
energy band 0.15-10keV. We extracted the data from the public archive
and processed the data using the {\xmm} SAS v6.5 in a similar manner
to that described in Ramsay et al (2005a). Data were barycentrically
corrected and in units of Terrestrial Time (TT).

\begin{table}
\begin{center}
\begin{tabular}{lrrr}
\hline
{\sl XMM} & Start Date & Duration & Mean EPIC\\
 Orbit   &            & (ksec)   & pn (Ct/s) \\
\hline
0880 & 2004-09-28 & 17.0 & 1.08 \\
0882 & 2004-10-02 & 19.0 & 1.16 \\
\hline
{\chan} & Start Date &  Duration & Mean {\chan} \\
Obs Id  &            & (ksec)    & Ct/s \\
\hline
 4519     & 2004-04-05 & 15.4 &   0.44 \\
 4520     & 2004-06-13 & 14.7 &   0.37 \\
 4521     & 2004-09-05 & 15.2 &   0.34 \\
\hline
\end{tabular}
\end{center}
\caption{The log of the observations of RX J1914+24 reported
here. Each observation was continuous with no data gaps.  Top panel:
The {\xmm} observations of RX J1914+24 made in 2004. The earlier
observations are detailed in Ramsay et al (2005a).  Bottom panel: The
{\chan} made in 2004. The earlier observations are detailed in
Strohmayer (2004b).}
\label{log}
\end{table}

\subsection{{\chan} observations}

Observations of RX J1914+24 made using {\chan} on 5 occasions were
reported by Strohmayer (2004b). Since then another 3 observations have
entered the archive. All of these observations were made in continuous
clocking mode: in this mode a one-dimensional image of the sky is
produced. We have extracted all of the available data from the archive
and corrected and analysed them in a similar manner to that carried
out by Strohmayer (2004b).

\section{Determining the rate of spin-up}
\label{spinup}

\begin{figure}
\begin{center}
\setlength{\unitlength}{1cm}
\begin{picture}(6,5.5)
\put(-2.8,-7){\includegraphics{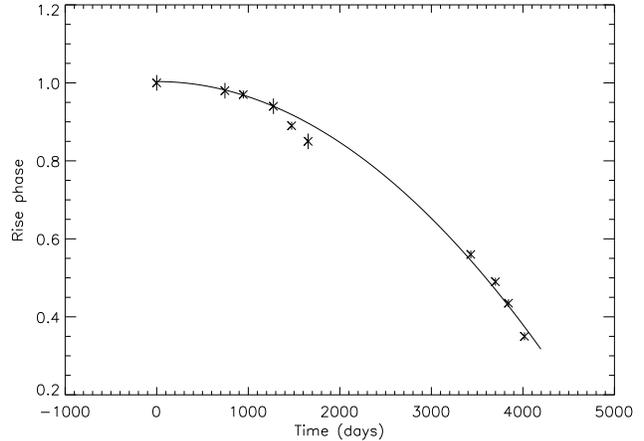}}
\end{picture}
\end{center}
\caption{The phase of the sharp rise to maximum X-ray flux determined
from {\ros}, {\chan} and {\xmm} data (see Ramsay et al 2005a for
details of these observations). The more recent point refers to the new
{\xmm} data presented here, with the preceding point refers to {\chan}
data taken in April 2004.}
\label{rise}
\end{figure}

Ramsay et al (2005a) folded all available X-ray data ({\ros}, {\asca},
{\chan} and {\xmm}) on the best fit initial period of Strohmayer
(2004b) (ie without the period derivative) and found that the sharp
rise of the X-ray flux arrives earlier in phase over time. A best fit
to the data showed that RX J1914+24 is spinning up at a rate of
3.2$\times10^{-12}$ s/s, which was similar to that reported by
Strohmayer (2002, 2004b). We did the same analysis as done in Ramsay et
al (2005a) and show the phase of the rise in X-ray flux for the {\sl
Chandra} data taken in April 2004 and the new {\xmm} data in Figure
\ref{rise}. We find a best fit of $\dot{P}=-3.31\pm0.09\times10^{-12}$
s/s (the error being determined using a bootstrap method), which is
consistent with the value we determined previously.

\section{Searching for periods other than the 569 sec period}

Strohmayer (2004b) detected sideband structure in the power spectrum of
RX J1914+24 in observations made using {\chan} but concluded that this
was likely to have been caused by the dithering pattern of those
observations. On the other hand, longer term variations have been seen
in X-ray data (Ramsay et al 2000, Strohmayer 2004b).

\begin{figure}
\begin{center}
\setlength{\unitlength}{1cm}
\begin{picture}(7,13)
\put(-1.,-0.4){\includegraphics{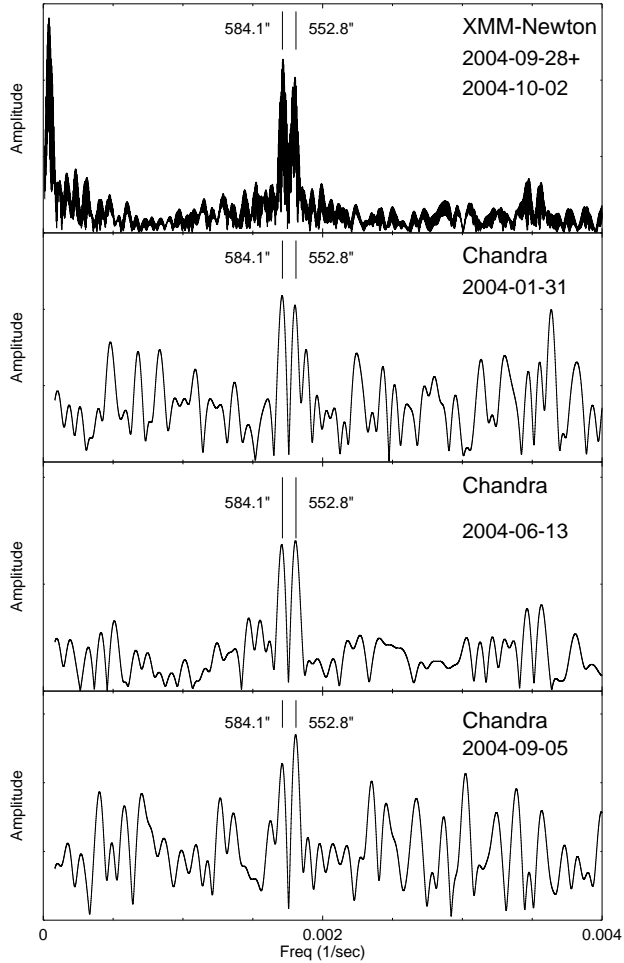}}
\end{picture}
\end{center}
\caption{The pre-whitened amplitude spectra of 5 sets of X-ray
data. They have been pre-whitened by subtracting the mean X-ray
profile appropriate to each epoch.}
\label{prewhite}
\end{figure}

We have searched for evidence of periods other than the 569 sec period
in these new X-ray data. We performed a Discrete Fourier Transform
(DFT) on the combined data taken in {\xmm} orbits 0718 and 0721 and
also orbits 0880 and 0882 as well as each of the {\chan}
observations. As expected, the 569 sec period and its harmonics were
seen strongly in all the amplitude spectra. To pre-whiten the data we
used two methods: we pre-whitened in the usual manner using a sinusoid
at predicted period at the observation epoch (based on the initial
period of Strohmayer (2004b) and the spin-up rate reported here). We
also pre-whitened the data by subtracting the mean X-ray pulse profile
appropriate to each individual observation. The latter is more
appropriate since the X-ray pulse profile is non-sinusoidal.  As
expected most of the peaks are removed after this procedure. However,
in the second epoch set of {\xmm} data (0880 and 0882), side-band
structure is clearly present at 584.1 sec and 552.8 sec in the
pre-whitened spectrum (Figure \ref{prewhite}). They are located in the
wings of the main peak in the original amplitude spectrum and have an
amplitude of $\sim$1/10 of the main peak. In the earlier {\xmm} data
(0718 and 0721) there is only weak evidence of these periods. In the
{\chan} dataset of 2004 June 13 we also find strong evidence for
periods very similar to those seen in the {\xmm} data (Figure
\ref{prewhite}). In two others, (2004 Jan 31 and 2004 Sept 5) there
are peaks in the amplitude spectra which correspond to these periods,
although they are only marginally significant compared to other peaks
in the spectra.  There is no evidence for these periods in the longest
{\chan} dataset which lasted 35 ksec. We also determined how the X-ray
pulse profile changed over the individual observation duration. We
find that the mean profile brightens over the duration of both the
0880 and the 0882 observations.
 
We believe that 553 and 584 sec periods could result when a modulation
signal (ie the 569 sec period) is mixed with another longer term
variation (either a periodic or non-periodic modulation). {\sl If} it
was a periodic modulation, then its period would be the difference
between the 569 sec period and the 553 and 584 sec periods: just over
6 hrs. Alternatively if a modulation signal is mixed with either a 553
sec or 584 sec period then this would give rise to the longer term
trend in the data. The question remains if this longer term trend is
periodic or due to the secular variability that have already been
observed in this system: we discuss this further in \S
\ref{discussion}.

\section{Spectra}

Ramsay et al (2005a) showed the X-ray spectrum of RX J1914+24 was
poorly modelled using an absorbed blackbody model. Their best fit was
obtained using an absorbed blackbody plus a broad Gaussian line near
0.6keV. This fit was still poor (\rchi=1.63) and the model was
unlikely to be physically realistic.

\begin{figure}
\begin{center}
\setlength{\unitlength}{1cm}
\begin{picture}(6,6)
\put(-2,-0.8){\includegraphics{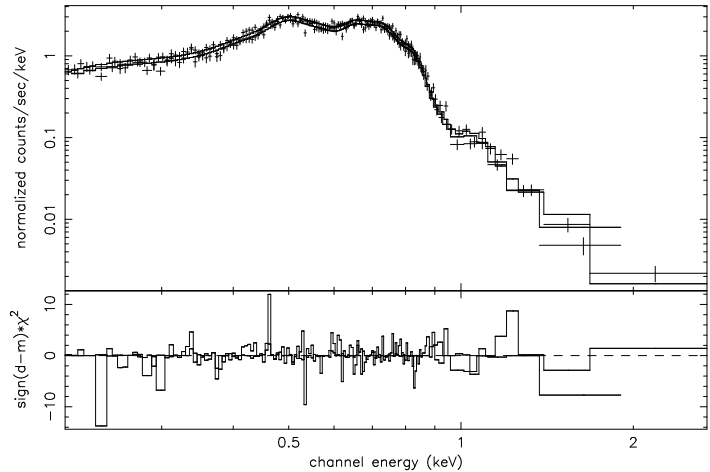}}
\end{picture}
\end{center}
\caption{The {\xmm} EPIC pn spectra from the data taken in 2004. The
solid line is the best fit to the model consisting of an absorbed thermal 
plasma model with an edge.}
\label{spec_fit}
\end{figure}

With the better quality data afforded by the new {\xmm} data we have
re-visited the X-ray spectrum of RX J1914+24. We fitted the EPIC-pn
data from {\xmm} orbits 0880 and 0882 separately. Given the unphysical
nature of the previous fits and working from the satisfactory fits to
our X-ray observations of other compact binaries (the AM CVn stars,
Ramsay et al 2005b) we tried fitting the spectra with an absorbed
thermal plasma model with variable elemental abundance ({\tt vmekal}
in XSPEC). This model still gave a poor fit to the data (\rchi=4.6 for
the 0880 data). However, when we added an edge to the model we
obtained a good fit to the 0880 data (\rchi=0.97, 122 dof; Figure
\ref{spec_fit}). The fit to the 0882 data using the same model was
slightly poorer (\rchi=1.21, 141 dof). Elements which were required in
the fit include N, O, Ne, Ca, Fe for the 0880 data, with some evidence
for Na in the 0882 data. The edge is located at 0.83keV which is
consistent with it being due to OVIII.  The absorption derived from
the fits is $\sim1.2\times10^{21}$ \pcmsq. We note we cannot obtain a
good fit with this model when we fix the absorption at
8.3$\times10^{21}$ \pcmsq, the larger value determined from optical
interstellar absorption features (Steeghs et al 2006).

The observed flux in the 0.1-10keV energy band (1.30$\times10^{-12}$
\ergscm for the orbit 0880 data and 1.48$\times10^{-12}$ \ergscm for
the orbit 0882 data) is very similar to that found using the blackbody
plus Gaussian line of Ramsay et al (2005a). However, the unabsorbed
bolometric flux in the fit presented here (1.14$\times10^{-11}$
\ergscm; 0880 and 1.24$\times10^{-11}$ \ergscm; 0882) is substantially
reduced compared to the earlier results (3.2$\times10^{-10}$ \ergscm)
where a blackbody spectrum was used.

\section{Discussion}

\subsection{The periods in RX J1914+24}
\label{discussion}

Until now there has been no convincing evidence for any other periods
in RX J1914+24 apart from the 569 sec period. In the case of 5 X-ray
observations we have found evidence for power at 552 and 584 sec. This
changes our perception of RX J1914+24. The data are clear - the X-ray
flux clearly increases in the {\xmm} data taken in both orbit 0880 and
in orbit 0882. This could be considered to be the result of an
underlying $\sim$6 hour period, or it could be the result of secular
variability known already in this system at longer timescales (Ramsay
et al 2000, Strohmayer 2004b). Both could produce the observed
amplitude spectrum.

\begin{figure}
\begin{center}
\setlength{\unitlength}{1cm}
\begin{picture}(6,6)
\put(-2,-0.3){\includegraphics{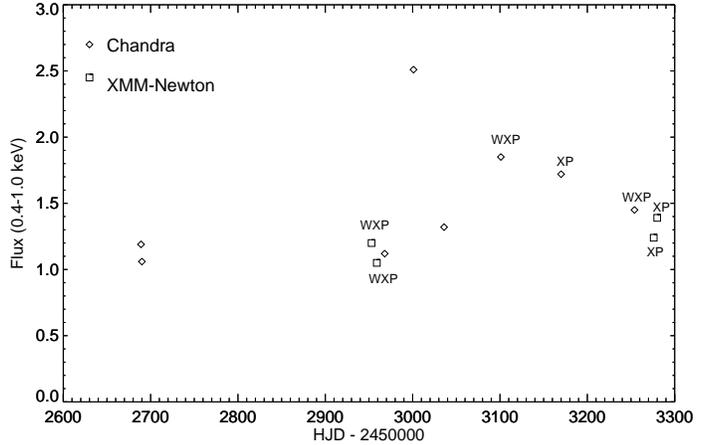}}
\end{picture}
\end{center}
\caption{The observed flux in the 0.4-1keV energy band (units of
$10^{-12}$ \ergscm) determined in {\chan} and {\xmm}
observations. `XP' denotes the presence of periods at 552 and 583 sec:
in the case of the most recent {\chan} observation the significance is
marginal - `WXP' weak X-ray periods.}
\label{long_term}
\end{figure}

To explore this question, we have obtained the mean X-ray flux in each
of the {\chan} and {\xmm} observations. Since the {\chan} ACIS-I
detector has a low effective area at energies less than $\sim$0.4keV,
we determined the observed mean flux in the 0.4-1.0keV energy band.
We show how the flux in this band varies over time in Figure
\ref{long_term}. We also show those epochs at which we detected the
552 sec and 584 sec periods which could be signatures of a longer
period trend in the observation. Although not conclusive, it appears
that at the higher flux levels there is no (or weak) evidence of these
other periods, so that shorter ($<$1 day) timescale variations may
take place only when the system is emitting at intermediate or low
levels. The detection of power at 552 and 584 sec does not appear to
be related to the observation length since the longest observation
(the {\chan} observation of Feb 2003 which lasted 35 ksec) did not
show obvious power at these periods.

At this stage, we cannot distinguish between the X-ray long term
behaviour being due to random changes in either the accretion rate (in
the case of the accreting models) or the electrical current
dissipation (in the case of the unipolar-inductor model). We urge the
search for radial velocity variations on a timescale of less than 1
day, or dedicated X-ray and/or optical observations of this system to
search for variations on this period.

{\it If} the longer term variation in the X-ray flux is periodic, then
we showed earlier that its period would approximately be 6 hrs.  Smith
\& Dhillon (1998) show that for a binary system with this orbital
period then its spectral type would be expected to be a mid K-type,
although the range is up to 5 sub-classes.  We note with interest the
discovery of Steeghs et al (2006) of features in the optical spectrum
of RX J1914+24 which appear to resemble a G9/K0 spectral
type. Therefore, an orbital period of 6 hrs would be marginally
consistent with this.

\subsection{The X-ray spectrum of RX J1914+24}

We have shown that the derived luminosity of RX J1914+24 is highly
model dependent. Further, the distance to RX J1914+24 is still rather
uncertain, but Steeghs et al (2006) suggest a distance of 1kpc based
on its optical spectrum. For a distance of 1kpc the mean bolometric
luminosity from the {\xmm} orbits 0880 and 0882 data is therefore
1.4$\times10^{33}$ \ergss - an order of magnitude lower than the
previous luminosity of 3.8$\times10^{34}$ \ergss. Ramsay et al (2005a)
went further and applied a correction to account for projection
affects giving a luminosity of $\sim1\times10^{35}$ \ergss. Since the
new model is an optically thin plasma, a correction effect for
projection is now no longer applicable. Moreover, absorption derived
from the X-ray fits is $\sim$15\% of that derived from the optical
data of Steeghs et al (2006) based on a G9 spectral type. If the G
star is not the secondary, then we can reduce the distance by a
similar factor and the luminosity is therefore 3$\times10^{31}$
\ergss. In this case, the X-ray luminosity of RX J1914+24 is more than
$\sim$3 orders of magnitude less than that found by Ramsay et al
(2005a).

In the unipolar-inductor model of Wu et al (2002) the amount of power
which is emitted is determined by various factors, with the degree of
asynchronisity being one of the key determinants (along with component
masses and magnetic field strength). However, to produce the observed
luminosity, the asynchronisity would have to be 0.001 or less (Wu et
al 2002). This reduces the objections to the UI model on energetic
grounds raised by Marsh \& Nelemans (2005) to a significant degree.

The abundances are clearly non-solar, which is consistent for an
evolved secondary as is seen in the fits to spectra of AM CVn systems
(eg Marsh, Horne \& Rosen 1991, Strohmayer 2004a, Ramsay et al
2005b). If this model reflects the properties of the material leaving
the secondary star then it rules out the G9/K0 star seen in the
optical spectrum being the secondary star.

\section{acknowledgments}

This is work based on observations obtained with {\xmm}, an ESA
science mission with instruments and contributions directly funded by
ESA Member States and the USA (NASA). The {\xmm} data were extracted
from the archive at Vilspa and the {\chan} data from the HEASARC
archive at Goddard Space Flight Center.

\end{document}